\documentclass[reprint, amsmath,amssymb, aps,pra,superscriptaddress]{revtex4-2}

\usepackage{graphicx}
\usepackage{dcolumn}
\usepackage{bm}
\usepackage[margin=2.0cm]{geometry}
\usepackage{amsmath}
\usepackage{braket}
\usepackage{appendix}
\usepackage{float}
\usepackage{siunitx}
\usepackage{amssymb}
\usepackage{color}
\usepackage{multirow}
\usepackage{longtable}
\usepackage{ulem}
\usepackage{soul}
\usepackage{adjustbox}
\usepackage{tabularx}

\begin{document}

\title{Moderate-terahertz-induced plateau expansion \\ of high-order harmonic generation to soft X-ray region}

\author{Doan-An Trieu}
    \email{trieudoanan@duytan.edu.vn}
    \affiliation{Institute of Fundamental and Applied Sciences, Duy Tan University, Ho Chi Minh City 700000, Vietnam}
    \affiliation{Faculty of Natural Sciences, Duy Tan University, Da Nang City 550000, Vietnam}
\author{Duong D. Hoang-Trong}
    \affiliation{Simulation in Materials Science Research Group, Science and Technology Advanced Institute, Van Lang University, Ho Chi Minh City, Vietnam}
    \affiliation{Faculty of Applied Technology, Van Lang School of Technology, Van Lang University, Ho Chi Minh City, Vietnam}
\author{Cam-Tu Le}
    \affiliation{Atomic Molecular and Optical Physics Research Group, Institute for Advanced Study in Technology, Ton Duc Thang University, Ho Chi Minh City 72912, Vietnam}
    \affiliation{Faculty of Applied Sciences, Ton Duc Thang University, Ho Chi Minh City 72912, Vietnam}
\author{Sang Ha}
    \affiliation{Computational Physics Key Laboratory K002, Department of Physics, Ho Chi Minh City University of Education, Ho Chi Minh City 72722, Vietnam} 
\author{Ngoc-Hung Phan}
    \affiliation{Computational Physics Key Laboratory K002, Department of Physics, Ho Chi Minh City University of Education, Ho Chi Minh City 72722, Vietnam} 
 \author{F. V. Potemkin} 
    \affiliation{Faculty of Physics, M. V. Lomonosov Moscow State University, Leninskie Gory 1/2, Moscow 119991, Russian Federation}
\author{Van-Hoang Le}
    \affiliation{Computational Physics Key Laboratory K002, Department of Physics, Ho Chi Minh City University of Education, Ho Chi Minh City 72722, Vietnam} 
\author{Ngoc-Loan Phan}
    \email{loanptn@hcmue.edu.vn (Corresponding author)}
    \affiliation{Computational Physics Key Laboratory K002, Department of Physics, Ho Chi Minh City University of Education, Ho Chi Minh City 72722, Vietnam}

\date{\today}

\begin{abstract}
Extending the high-harmonic cutoff with experimentally accessible fields is essential for advancing tabletop coherent extreme-ultraviolet (EUV) and soft X-ray sources. Although terahertz (THz) assistance offers a promising route, cutoff extension at weak, laboratory-accessible THz strengths remains poorly understood. In this Letter, we investigate THz-assisted high-order harmonic generation (HHG) using time-dependent Schr\"{o}dinger equation simulations supported by classical analysis and Bohmian-based quantum dynamics. By mapping the plateau evolution versus THz field strength, we demonstrate that even weak THz fields extend the cutoff, producing a pronounced ``fish-fin'' structure whose dominant branches saturate near $I_p + 8U_p$. We attribute this extension to long electron excursions spanning multiple optical cycles before recombination. Our results establish that this cutoff-extension mechanism is robust across atomic species and driving parameters, indicating that laboratory-scale THz fields enable practical cutoff control and access to high-energy HHG.
\end{abstract}

\maketitle

\textit{Introduction.} High-order harmonic generation (HHG) is a key tabletop source of coherent extreme-ultraviolet (EUV) and X-ray radiation, enabling advances in strong-field physics and attosecond science~\cite{Chergui:NatRevPhys23,Popmintchev:pnas09,Weissenbilder:NatRevPhys22,Borrego-Varillas:RepProPhys22}.  Although HHG from condensed media has recently been explored, gas-phase HHG remains the most effective route for high-energy coherent emission due to its higher attainable cutoff energies~\cite{Vampa:jpb17,Luu:NatCom18}. The conventional gas-phase cutoff, $I_p + 3.17U_p$ (where $I_p$ is the ionization potential and $U_p$ the ponderomotive energy), follows from the three-step model and sets the standard HHG energy limit~\cite{Corkum:prl93,Lewenstein:pra94}.

Over the past two decades, substantial efforts have focused on extending the HHG cutoff, with field engineering emerging as an effective route to control electron trajectories~\cite{Bandrauk:PRA97,Carrera:PRA07,Raab:RevSciInstru24,Yuan:PRA11,Ge:OE15,Hong:JPB07,Hong:OE09}. A particularly simple strategy is to superimpose a static electric field on the driving infrared (IR) pulse, which can theoretically extend the cutoff to $I_p + 9.1U_p$~\cite{Hong:JPB07,Hong:OE09,Taranukhin:josab00,Odzak:pra05,Rumiantsev:pra25,Yuan:PRA11,Koushki:CP21,Ge:OE15,Ge:LPL20,Mohebbi:aplB,Wang:jpb98}. However, achieving this extension requires static fields of about $39\%$ of the fundamental field (on the order of $100~\mathrm{MV/cm}$)~\cite{Hong:JPB07,Hong:OE09,Taranukhin:josab00,Odzak:pra05,Yuan:PRA11,Ge:OE15,Koushki:CP21,Rumiantsev:pra25}, which remains experimentally challenging~\cite{Fulop:ThzRev20,Choi:JMCC24,Rumiantsev:pra25}.

Recent advances in terahertz (THz) technology have enabled THz fields to serve as dynamic substitutes for static biasing in HHG~\cite{Li:NC23,Rumiantsev:pra25}, opening new opportunities to retrieve target structural dynamics and THz-field temporal profiles~\cite{Li:pra20,Li:NC23,Rumiantsev:pra25,Liu:pra25,Trieu:pra23,Trieu2:pra24,Trieu:pra24}. In typical THz-assisted HHG experiments, THz amplitudes are only a few percent of the driving IR field and are readily generated with tabletop sources~\cite{Li:NC23,Shulyndin:JETP25}. However, significant THz-assisted cutoff extension has so far required THz fields of about $100~\mathrm{MV/cm}$~\cite{Hong:JPB07,Rumiantsev:pra25,Odzak:pra05,Hong:OE09,Taranukhin:josab00}, accessible only in specialized facilities due to conversion inefficiencies and damage constraints~\cite{Fulop:ThzRev20}. Extending the HHG cutoff with weak, laboratory-scale THz fields therefore remains a key challenge.

In addition, weak static electric fields have been shown to induce multiple HHG plateaus, indicating cutoff extensions~\cite{Taranukhin:josab00,Odzak:pra05,Hong:JPB07}. This suggests that experimentally accessible weak THz fields may also enable cutoff extension. However, no clear rule exists for how weak THz fields govern cutoff extension, nor whether the resulting plateaus occur only at specific field strengths or persist over a broad range. This motivates a systematic investigation of THz-dependent cutoff behavior.

\begin{figure*}[htb!]
 	\centering
\includegraphics[width=1.\linewidth,clip=true]{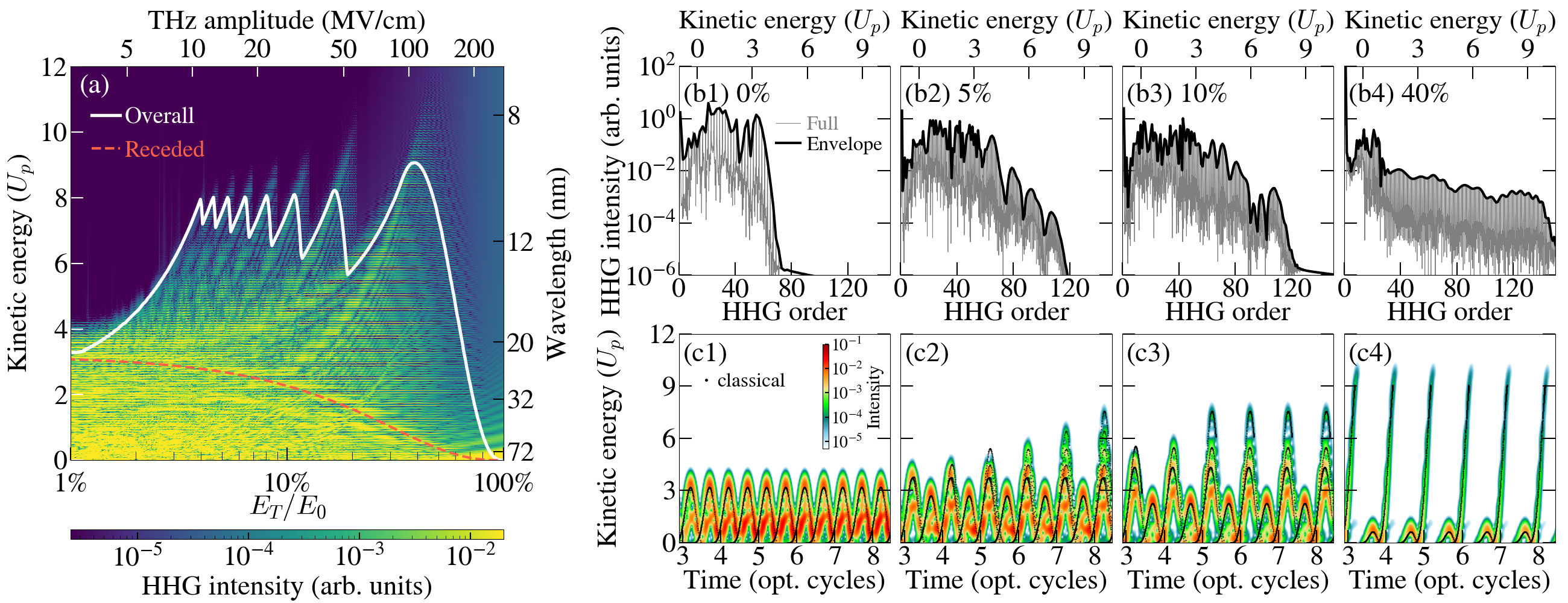}
 	\caption{(a) HHG from a hydrogen atom in the combined IR-THz fields with varying THz strengths, where the color encodes the HHG intensity. Panels [(b1)-(b4)] show representative HHG spectra, and panels [(c1)-(c4)] display the corresponding time-frequency profiles for selected THz fields. In panels [(b1)-(b4)], the gray curves represent the HHG spectra and the solid black curves indicate their spectral envelopes. In panels [(c1)-(c4)], the dotted curves show the classical returning-electron kinetic energies. In panel (a), the overall HHG cutoff exhibits a characteristic ``fish-fin'' structure consistent with classical simulations (white solid curve), indicating efficient EUV and soft-X-ray generation at moderate THz strengths. For $\alpha$ in the range of $4\%$-$40\%$, a multiplateau structure emerges due to the imbalance between adjacent attosecond bursts [(c2)-(c4)], where one burst group forms the first plateau [red dashed curve in panel (a)] and the other contributes to its extension.
    } 
    \label{fig:compre}
\end{figure*}

In this Letter, we systematically map the HHG plateau evolution as a function of THz-field strength in combined IR–THz fields. By numerically solving the time-dependent Schr\"{o}dinger equation (TDSE) for representative atomic targets (H, He, Ne, Ar), we uncover a universal THz-assisted cutoff-extension rule that manifests as a characteristic ``fish-fin'' structure. This ``fish-fin'' structure establishes a practical rule indicating that substantial cutoff extension can be achievable with weak, laboratory-scale THz fields. To elucidate the underlying mechanism, we perform classical and Bohmian-based trajectory analysis, revealing that the cutoff extension originates from long electron excursions spanning multiple optical cycles.

\textit{THz-assisted HHG: universal ``fish-fin'' structure.} HHG from an atom under varying THz fields is computed by numerically solving the TDSE, written in atomic units (a.u.) as
\begin{equation}
i \dfrac{\partial}{\partial t}\psi(\mathbf{r},t) = \left[ -\dfrac{1}{2}\nabla^2 + V_\mathrm{c}(\mathbf{r}) +  \mathbf{r} \cdot \mathbf{E}(t) \right]\psi(\mathbf{r},t), \label{eq:tdse}
\end{equation}
Here, $\mathbf{E}(t)$ is the combined field polarized along the $x$ axis, with the amplitude given by $E(t) = E_0 f(t) \cos(\omega_0 t) + E_T\cos(\omega_T t)$, defined within the time interval $ [-\tau/2,\tau/2]$, where $\tau$ is the IR pulse duration. Here, $E_0$ and $\omega_0$ denote the IR peak field and carrier frequency, while $E_T$ and $\omega_T$ correspond to the THz field. $V_\mathrm{c}(\mathbf{r})$ represents the atomic potential. Because the IR and THz fields are collinearly polarized, electron dynamics are predominantly along the polarization axis, motivating the use of a reduced-dimensional model~\cite{Chirila:pra10,Majorosi:pra18,Trieu:pra23}. We therefore employ a density-based one-dimensional soft-Coulomb potential, which yields HHG spectra in good agreement with reported three-dimensional calculations~\cite{Majorosi:pra18}. We further validate this model by benchmarking THz-assisted HHG spectra against three-dimensional TDSE simulations at representative THz field strengths, finding good agreement~\cite{Trieu:pra23}. Additional numerical details are given in the Supplemental Material~\cite{suppl}.

Figure~\ref{fig:compre}(a) shows HHG spectra from hydrogen as a function of THz field strength. For convenience, the left axis indicates the kinetic energy of the returning electron $K$, corresponding to the harmonic photon energy $I_p + K$. The THz strength is parameterized by $\alpha = E_T/E_0$. The driving IR pulse consists of 12 cycles with a flat-top envelope (four-cycle turn-on and turn-off), an intensity of $10^{14}$~W/cm$^2$, and a wavelength of 1200~nm, while a THz field with a frequency of 1.3~THz (231~$\mu$m) is used. These parameters are chosen as representative values for visualization.

The HHG spectra in Fig.~\ref{fig:compre}(a) exhibit a plateau followed by a sharp cutoff for all values of $\alpha$. However, the overall cutoff does not vary smoothly with $\alpha$; instead, it forms a distinct \textit{``fish-fin'' structure} with pronounced rays. As the THz strength increases, the cutoff extends from the conventional limit $I_p + 3.17U_p$ (with $U_p = E_0^2/4\omega_0^2$) to approximately $I_p + 8.0U_p$, reaching the EUV and soft–X-ray regime at $\alpha = 4\%$ ($E_T \approx 12$~MV/cm). Over a broad range of $4\% \le \alpha \le 40\%$, the cutoff fluctuates within an extended energy window. At $\alpha = 40\%$ ($E_T \approx 100$~MV/cm), the cutoff attains a maximum of $I_p + 9.1U_p$, consistent with previous reports~\cite{Hong:JPB07,Hong:OE09,Taranukhin:josab00,Odzak:pra05,Rumiantsev:pra25}, before decreasing at higher $\alpha$. This ``fish-fin'' behavior indicates that efficient EUV and soft–X-ray HHG can be achieved with moderate, laboratory-scale THz fields.

In addition to cutoff extension, Fig.~\ref{fig:compre}(a) reveals a pronounced \textit{multiplateau structure} emerging for \mbox{$\alpha \gtrsim 3\%$}. Beyond the primary plateau (red dashed curve), higher plateaus follow the rays of the ``fish-fin'' pattern. Although similar multiplateau structures were reported previously~\cite{Odzak:pra05,Taranukhin:josab00}, we demonstrate that both the number and morphology of the plateaus depend sensitively on $\alpha$, as illustrated for $\alpha = 5\%$ and $10\%$ in Figs.~\ref{fig:compre}[(b2)-(b3)]. For \mbox{$\alpha \gtrsim 10\%$}, the number of plateaus decreases [Fig.~\ref{fig:compre}(b4)]. Beyond these structural changes, the cutoff behavior exhibits a striking contrast: as $\alpha$ increases, the first-plateau cutoff (red dashed curve) shifts to lower harmonic orders, whereas the overall cutoff (white solid curve) continues to extend with pronounced fluctuations. As shown in Figs.~\ref{fig:compre}[(b1)–(b4)], the first-plateau cutoff decreases from harmonic order 55 to 21 as $\alpha$ increases from 0 to 40\%, while the overall cutoff extends from 55 to 137.

We further verify the ``fish-fin'' structure and associated cutoff extension in other atomic targets driven by combined IR–THz fields. Calculations for rare-gas atoms He, Ne, and Ar (not shown) reveal the same characteristic pattern, demonstrating that the ``fish-fin'' THz-induced cutoff extension is robust across atomic species. We also vary the IR intensity and wavelength and consistently recover the ``fish-fin'' structure, indicating robustness with respect to driving-field parameters. The IR intensity is chosen to allow efficient ionization while avoiding saturation and excessive ground-state depletion~\cite{Keldysh:JETP64,Brabec:RevModPhys00,Phan:josab20}. These results establish the ``fish-fin'' structure as a \textit{universal feature} of atomic THz-assisted HHG that is experimentally accessible.

Our observation of the ``fish-fin'' structure demonstrates that coherent EUV and soft X-ray harmonics can be generated using moderate THz fields of only $10-$tens of $\mathrm{MV/cm}$, rather than the extremely strong fields $\sim 100~\mathrm{MV/cm}$ suggested previously \cite{Hong:JPB07,Hong:OE09,Taranukhin:josab00,Odzak:pra05,Rumiantsev:pra25}. Crucially, THz fields in this range are readily achievable with standard tabletop laser-driven sources, such as optical rectification in organic nonlinear crystals and air-plasma-based schemes \cite{Rumiantsev:pra25,Li:NC23,Shulyndin:JETP25,Fulop:ThzRev20,Choi:JMCC24}. In contrast, the generation of quasi-stationary THz fields approaching $100~\mathrm{MV/cm}$ remains technologically challenging and typically requires either large-scale accelerator-based facilities~\cite{Carr:2002,Xu:NatPho21}, or sophisticated multi-crystal schemes relying on wide-aperture organic crystals combined with precise coherent field interference. Even reaching field on the order of $50~\mathrm{MV/cm}$ is highly nontrivial and has so far been demonstrated only in a handful of specialized laboratories worldwide. For example, achieving such field strengths requires a faceted assembly of four organic crystals pumped at $1.2$--$1.5~\mu\mathrm{m}$ with pulse energies of $\sim 30~\mathrm{mJ}$ and fluences approaching $10~\mathrm{mJ/cm^2}$ \cite{Vicario:ol14}, which can be realized using Cr:Forsterite-based laser systems \cite{Pushkin:Photonics}. Therefore, the observation of the ``fish-fin'' structure demonstrates that coherent soft X-ray harmonics can be accessed using experimentally realistic tabletop THz field strengths, substantially relaxing the requirement for ultra-high THz fields previously considered necessary.

\textit{THz-dependent cutoff energies of emission bursts.} To gain deeper insight into the origin of the ``fish-fin'' structure, we compute harmonic time–frequency (TF) profiles using a Gabor transform~\cite{Chirila:pra10} for different THz field strengths [Figs.~\ref{fig:compre}(c1)–(c4)]. These profiles reveal trains of attosecond bursts separated by half an optical cycle, whose temporal structure depends sensitively on the THz field. In the absence of THz assistance, adjacent bursts are identical owing to the inversion symmetry of the atomic target combined with a half-cycle time translation of the field [Fig.~\ref{fig:compre}(c1)]. In contrast, the THz field breaks this symmetry, producing an alternating burst pattern [Figs.~\ref{fig:compre}(c2)–(c4)]. In these cases, the bursts can be categorized into two groups: the first corresponds to emission around $(3.2 + k)T_0$, and the second around $(3.7 + k)T_0$, where $T_0 = 2\pi/\omega_0$ is the optical period and $k = 0-8$ for the used 12-cycle IR pulse.

Figures~\ref{fig:compre}[(c1)-(c4)] show that, for each value of $\alpha$, the bursts in the second group are nearly identical, and their maximum photon energies are lower than those of the first group. As a result, these bursts form the first plateau, with its cutoff determined by their maximum energy. Furthermore, as the THz field strength increases, their cutoff energies decrease rapidly, explaining the recession of the first-plateau cutoff in the HHG spectrum, as indicated by the red dashed curve in Fig.~\ref{fig:compre}(a).

By contrast, the attosecond bursts in the first group exhibit a pronounced THz dependence. For a weak THz field, e.g., $\alpha = 5\%$ in Fig.~\ref{fig:compre}(c2), the cutoff energies of these bursts increase stepwise in time, producing a multiplateau structure. For a stronger field, such as $\alpha = 10\%$ in Fig.~\ref{fig:compre}(c3), the cutoffs of bursts emitted in the latter half of the pulse become nearly identical, thereby reducing the number of plateaus. At an even higher field, $\alpha = 40\%$ [Fig.~\ref{fig:compre}(c4)], the bursts recover uniformity, leading to the emergence of a broadband second plateau~\cite{Hong:JPB07,Hong:OE09,Rumiantsev:pra25,Odzak:pra05,Taranukhin:josab00,Wang:jpb98}. %This broadband second plateau, characterized by nearly in-phase harmonics, is favorable for generating attosecond pulse trains~\cite{Hong:JPB07,Hong:OE09,Rumiantsev:pra25}.

In summary, the THz field reshapes the cutoff energies of attosecond bursts, thereby controlling the formation, suppression, and disappearance of multiple plateaus in the HHG spectrum. This establishes a direct link between the HHG spectral structure and attosecond burst emission, which is governed by the electron dynamics in the combined IR and THz fields, thus motivating a detailed trajectory-based analysis.

\begin{figure*}[t]
 	\centering
 	\includegraphics[width=0.9\linewidth,clip=true]{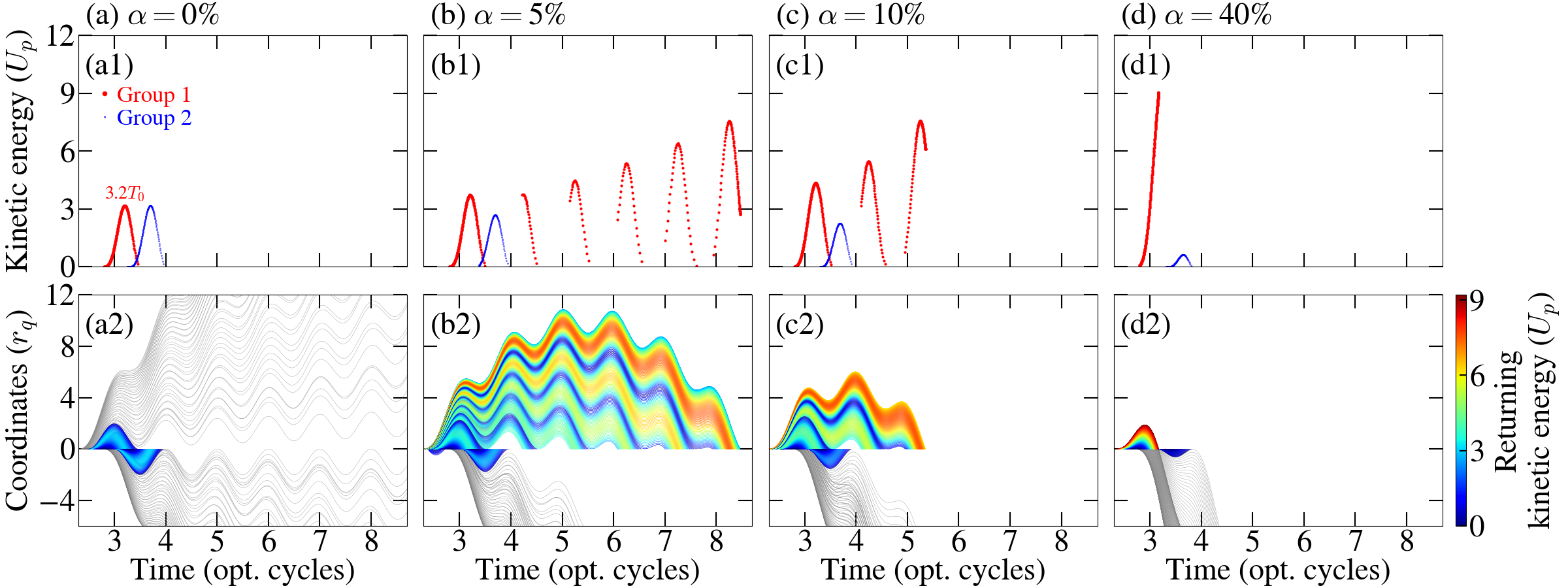}
 	\caption{Classical simulation of recombination kinetic-energy (first row) and electron excursion (second row) for different THz fields with $\alpha = 0$, 5\%, 10\%, and 40\%. For clarity, only trajectories ionized within one optical cycle are shown. In the first row, the blue and red curves correspond to the two attosecond-burst groups in Figs.~\ref{fig:compre}[(c1)-(c4)]). In the second row, gray curves denote all photoelectron trajectories, while colored curves highlight the returning ones whose colors encode their recombination kinetic energy. The IR-THz fields' parameters are the same as in Fig.~\ref{fig:compre}, except for the continuous IR. A ladder of long-traveling trajectories sets discrete burst cutoffs and underpins multiplateau HHG under moderate THz fields.}
    \label{fig1} 
 \end{figure*}

\textit{Electron-trajectory analysis: roles of long-traveling trajectories and kinetic-energy saturation.} Although the multiplateau structure has been qualitatively explained for specific THz field strengths using classical electron trajectories~\cite{Hong:JPB07,Hong:OE09,Odzak:pra05,Taranukhin:josab00,Wang:jpb98,Trieu:pra23,Trieu:pra24,Trieu2:pra24,Rumiantsev:pra25}, the physical origin of the ``fish-fin'' feature remains unresolved. Here, we identify electron trajectories to uncover the microscopic mechanism underlying this structure.

We simulate the classical motion of an ionized electron in combined IR and THz fields by solving Newton’s equation $\ddot{x}(t)=-E(t)$ with initial conditions $x(t_i)=\dot{x}(t_i)=0$, where $t_i$ is the ionization time~\cite{Corkum:prl93,Paulus:jpb94}. Sampling different ionization times yields an ensemble of trajectories. Most electrons drift away as photoelectrons, whereas a subset returns to the parent ion at $t_r$ satisfying $x(t_r)=0$. These returning electrons can recombine with the parent ion and emit harmonics with energy $\Omega = I_p + K$, where the kinetic energy at return is $K=\dot{x}^2(t_r)/2$. The IR field is treated as a continuous wave [$f(t)=1$] for simplicity. Besides, we also computed quantum Bohmian trajectories~\cite{Le:pra23,Le:pra24} and obtained consistent conclusions, so only the classical results are presented here for clarity.

Figure~\ref{fig1} shows the kinetic energy accumulated by returning electrons at recombination (first row) and their corresponding trajectories (second row) for $\alpha = 0,$ $5\%$, $10\%,$ and $40\%$. Owing to field periodicity, only electrons released within the first optical cycle are displayed. Figure~\ref{fig1}(a) shows that in the THz-free case ($\alpha = 0$), the kinetic energies in the two half cycles are identical due to symmetric electron propagation on both sides of the hydrogen atom. As indicated by the gray curves, electrons that travel beyond approximately $2r_q$ (with $r_q = E_0/\omega_0^2$ the quiver amplitude) cannot return to the parent ion. Consequently, the maximum return energy is limited to $3.17U_p$~\cite{Corkum:prl93,Lewenstein:pra94}.

However, the inclusion of the THz field breaks left–right symmetry and selectively favors long-traveling return trajectories. For $\alpha = 5\%$ [Fig.~\ref{fig1}(b2)], electrons launched into the $x<0$ side can return only if they remain close to the ionic core, which reduces the maximum recombination kinetic energy to $2.69 U_p$. The corresponding harmonics are emitted around $(3.7+k)T_0$ (with $k=0-8$) [Fig.~\ref{fig1}(b1)], forming the second group of attosecond bursts in Fig.~\ref{fig:compre}(c2) and giving rise to the first HHG plateau in Figs.~\ref{fig:compre}[(a),(b2)]. By contrast, in the $x>0$ region, all liberated electrons are driven back to the core, including those displaced beyond $2r_q$, with excursion times of up to six optical cycles. The resulting spread in travel times leads to distinct maximum kinetic energies at recombination near $(3.2+k)T_0$~\cite{Odzak:pra05,Taranukhin:josab00}, as shown in Fig.~\ref{fig1}(b1). This produces the stepwise attosecond bursts of the first group in Fig.~\ref{fig:compre}(c2), explaining the emergence of the multiplateau structure in the HHG spectra [Figs.~\ref{fig:compre}[(a),(b2)]]. Specifically, under this weak THz field, long-traveling trajectories can reach kinetic energies up to $\sim 8U_p$, generating high-energy harmonics extending into the soft-X-ray region.

\begin{table}[t]
\centering
\caption{Maximum kinetic energy $K_{\max}$ forming each ``fin ray'' at THz strength $\alpha$, with associated excursion time $t_e$ and maximum displacement $A_m$ obtained from classical simulation. Categories label successive fin rays. Field parameters are the same as in Fig.~\ref{fig1}.}
\label{tab:kine}
\begin{tabular}{c|ccccccccccccccccccccc}
%\hline
 Category & 1 & 2 & 3 & 4 & 5 & 6 & 7 & 8 & 9 & 10 \\
\hline \hline
$\alpha$ (\%)               & 38.6 & 16.5 & 10.8 & 8.0 & 6.4 & 5.3 & 4.6 & 4.0 & 3.5 & 3.2  \\
$K_{\max}$ ($U_{p}$) & 9.1 & 8.2 & 8.1 & 8.1 & 8.0 & 8.0 & 8.0 & 8.0 & 8.0 & 8.0 \\
$t_e$ ($T_0$) & 0.86 & 1.95 & 2.96 & 3.98 & 4.98 & 5.98 & 6.98 & 7.98 & 8.99 & 9.98  \\
$A_m$ ($r_q$)                 & 2.0 & 4.0 & 5.5 & 7.2 & 8.7 & 10.3 & 11.9 & 13.5 & 15.1 & 16.6  \\
\end{tabular}
\end{table}

\begin{figure}[htb!]
 	\centering
 	\includegraphics[width=.9\linewidth,clip=true]{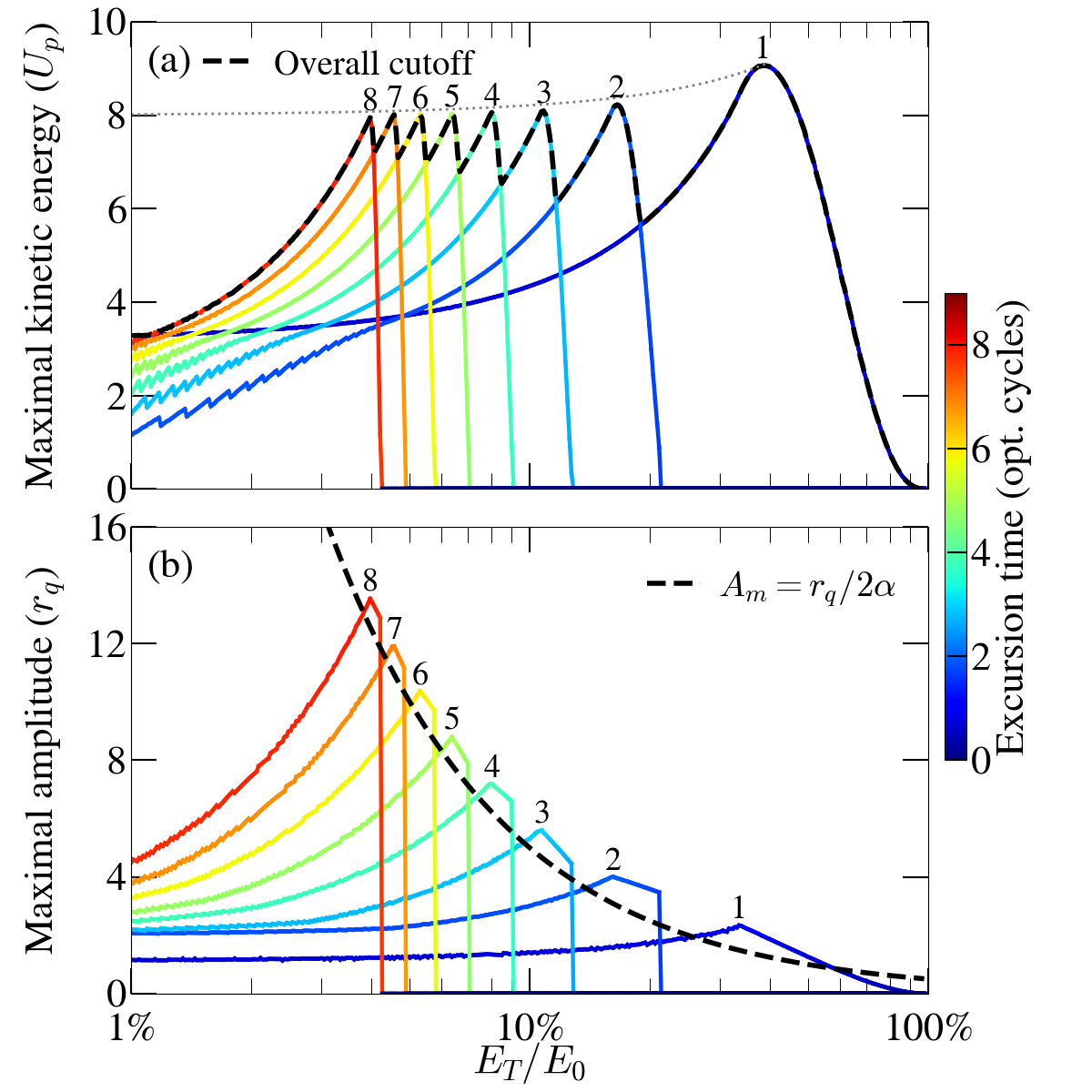}
 	\caption{Maximum kinetic energy (a) and maximum displacement (b) versus THz field strength for different electron excursion times (color-coded). In Panel (a), the black dashed curve marks the overall cutoff exhibiting the characteristic ``fish-fin'' structure, while the black dotted curve shows the trend of the maximum kinetic energy. In Panel (b), the dashed black curve marks the upper limit of returning-electron displacement. As the THz field weakens, longer excursions extend electron displacement while the return kinetic energy saturates near $8.0U_{p}$. 
} 
    \label{fig:class}
\end{figure}

As the THz strength increases (Fig.~\ref{fig1}[(c)-(d)]), the overall dynamics remain similar to the $\alpha=5\%$ case, but the maximum excursion of returning electrons moves closer to the parent ion. This shortens excursion times and, consequently, reduces the number of plateaus in the HHG spectra. At $\alpha=10\%$, the longest travel time decreases to about three optical cycles, with a maximum kinetic energy of $\sim 7.6U_p$. At $\alpha=40\%$, the strong THz field rapidly drives electrons back from the $x>0$ region while strongly suppressing returns from the $x<0$ side, yielding a maximum kinetic energy of $\sim 9.1U_p$, consistent with previous reports~\cite{Hong:JPB07,Hong:OE09,Rumiantsev:pra25,Odzak:pra05,Taranukhin:josab00,Wang:jpb98}.

The most intriguing unresolved issue is how the THz field governs the maximum kinetic energy gained by the THz-reshaped electron dynamics. To address this, Fig.~\ref{fig:class}(a) shows the dependence of the maximum kinetic energy on the THz field strength, while Fig.~\ref{fig:class}(b) presents the corresponding maximum electron displacement. 
The associated electron excursion time is encoded by the color scale.

The dashed black curve in Fig.~\ref{fig:class}(a) displays a distinct ``fish-fin'' pattern, repeatedly approaching about $8U_\mathrm{p}$ before attaining the highest value of $9.1U_\mathrm{p}$ at $\alpha = 39\%$. This structure originates from a ladder of long-traveling trajectory classes, where each successive “fin ray” (labeled from 1 to 8 in Figs.~\ref{fig:class}[(a), (b)]) corresponds to roughly one additional optical cycle in excursion time and a progressively larger spatial displacement. Remarkably, despite increasing excursion length and duration, the maximum return kinetic energy rapidly saturates near $8U_\mathrm{p}$. Simulations using longer driving pulses confirm that this saturation persists even at much lower THz field strengths, albeit at the expense of substantially larger electron excursions. Table~\ref{tab:kine} lists the $\alpha$ values at which local energy maxima occur and their associated excursion times. These THz-reshaped electron dynamics quantitatively reproduce the fish-fin-shaped cutoff (white solid curve) and the multiplateau structure in the TDSE HHG spectra (Fig.~\ref{fig:compre}), demonstrating excellent agreement with numerical results.

\textit{Analytical derivation of kinetic-energy saturation.} Following the classical equation for the electron motion in the combined field $E(t) \approx E_0 \cos (\omega_0 t) +E_T$, the return condition reads as $ \dfrac{E_0}{\omega_0^2}(\cos \omega_0 t_r - \cos\omega_0 t_i) + \dfrac{E_0}{\omega_0} t_e \sin\omega_0 t_i - \dfrac{E_T}{2}t_e^2= 0$, where $t_e = t_r - t_i$ is the excursion time. For THz-induced long-traveling trajectories $t_e\gg 2\pi/\omega_0$, this condition simplifies to $\dfrac{E_T}{2}t_e^2 - \dfrac{E_0}{\omega}t_e\sin\omega_0 t_i \approx 0$. Here, the THz field (first term) provides a steady outward drift, while the oscillating IR field (second term) supplies the restoring force that enables recollision. Their competition sets the maximum excursion time, yielding
  \begin{equation}
      t_e \approx \dfrac{T_0 }{\pi\alpha}\sin \omega_0 t_i
      \le \dfrac{T_0 }{\pi\alpha}.
  \end{equation}
The upper limit $t_e^\mathrm{max} = \dfrac{T_0 }{\pi\alpha}$ agrees well with the maximum excursion times obtained numerically and listed in Tab.~\ref{tab:kine}. Because long-traveling return trajectories are approximately symmetric, the electron reaches its largest displacement at mid-excursion $t_e/2$. The maximum displacement can therefore be estimated as
\begin{equation}
     A \approx \dfrac{r_q}{2\alpha} \sin^2 \omega t_i \le \dfrac{r_q}{2\alpha}, 
 \end{equation}
yielding an upper limit $A_m = \dfrac{r_q}{2\alpha}$, shown in Fig.~\ref{fig:class}(b) and consistent with numerical simulations. These relations establish a THz-controlled scaling law in which decreasing $\alpha$ simultaneously extends excursion time and spatial reach while maintaining recollision.

Within this long-traveling approximation, the electron velocity is predominantly gained from the THz field. Consequently, the velocity at the recombination instant is
\begin{equation}
      |v(t_r)| \approx E_T t_e \le \dfrac{2E_0}{\omega_0}.
\end{equation}
This imposes an upper bound on the return kinetic energy $K \le 8U_p$, establishing a saturated value of $8U_p$, in excellent agreement with the numerical simulations.

\textit{Conclusions.} In summary, we have identified a ``fish-fin'' cutoff structure in THz-assisted HHG that originates from long-traveling electron trajectories sustained over multiple optical cycles. Using TDSE simulations supported by classical and Bohmian-based trajectory analyses, we demonstrate that the cutoff evolves from $I_p +9.1 U_p$ toward a universal saturation at $I_p +8 U_p$ as the THz strength decreases. By analyzing THz-reshaped electron dynamics, we established a simple and physically transparent rule linking the attainable cutoff to the maximum excursion time, revealing kinetic-energy saturation as a fundamental constraint of long-trajectory dynamics. We further showed that this mechanism is broadly applicable to atomic systems below the saturation ionization regime. 

These findings demonstrate that cutoff control with laboratory level THz fields is feasible and that the ``fish-fin'' structure serves as a sensitive dynamical fingerprint of electron motion. The derived rule offers practical and predictive guidance for generating coherent EUV and soft X-ray emission using moderate THz fields available in current laboratories.

\textit{Acknowledgments.} This research is funded by the Vietnam National Foundation for Science and Technology Development (NAFOSTED) under Grant No. 103.01-2023.138 and carried out by the high-performance computing cluster at Ho Chi Minh City University of Education, Vietnam. The practice of experimental realization of THz fields was supported by the Russian Science Foundation (RSF), Project No. 25-22-00084.

N.-L. P. and D.-A. T. conceptualized the work, developed the methodology, and carried out the main analytical derivations. D.-A. T., C.-T. L., and N.-L. P. implemented the numerical simulations. D.-A. T., D. D. H.-T., S. H., and N.-H. P. validated the data. F. V. P. provided discussions on experimental feasibility. D.-A. T., N.-L. P., and V.-H. L. contributed to the interpretation of the physical mechanisms. N.-L. P. and V.-H. L. supervised the project. N.-L. P. and D.-A. T. wrote the original draft and finalized the manuscript. All authors discussed the results and contributed to reviewing and editing the manuscript.

\bibliographystyle{apsrev4-1}
\bibliography{refs.bib}

@misc{suppl,
howpublished="See Supplemental Materials for numerical method for solving time-dependent Schr{\"o}dinger equation and atomic potential models.",
}

@article{Xu:NatPho21,
  title={Cascaded high-gradient terahertz-driven acceleration of relativistic electron beams},
  author={Xu, Hanxun and Yan, Lixin and Du, Yingchao and Huang, Wenhui and Tian, Qili and Li, Renkai and Liang, Yifan and Gu, Shaohong and Shi, Jiaru and Tang, Chuanxiang},
  journal={Nat. Photonics},
  volume={15},
  number={6},
  pages={426--430},
  year={2021},
  publisher={Nature Publishing Group UK London}
}

@article{Carr:2002,
  author  = {Carr, G. L. and Martin, M. C. and McKinney, W. R. and Jordan, K. and Neil, G. R. and Williams, G. P.},
  title   = {High-power terahertz radiation from relativistic electrons},
  journal = {Nature},
  volume  = {420},
  pages   = {153--156},
  year    = {2002}
}

@Article{Pushkin:Photonics,
AUTHOR = {Pushkin, Andrey and Migal, Ekaterina and Suleimanova, Dina and Mareev, Evgeniy and Potemkin, Fedor},
TITLE = {High-Power Solid-State Near- and Mid-{I}{R} Ultrafast Laser Sources for Strong-Field Science},
JOURNAL = {Photonics},
VOLUME = {9},
YEAR = {2022},
NUMBER = {2},
ARTICLE-NUMBER = {90},
ISSN = {2304-6732},
}

@article{Vicario:ol14,
author = {C. Vicario and A. V. Ovchinnikov and S. I. Ashitkov and M. B. Agranat and V. E. Fortov and C. P. Hauri},
journal = {Opt. Lett.},
number = {23},
pages = {6632--6635},
publisher = {Optica Publishing Group},
title = {Generation of 0.9-mJ THz pulses in DSTMS pumped by a Cr:Mg2SiO4 laser},
volume = {39},
month = {Dec},
year = {2014},
url = {https://opg.optica.org/ol/abstract.cfm?URI=ol-39-23-6632},
doi = {10.1364/OL.39.006632},
}

@article{Brabec:RevModPhys00,
  title = {Intense few-cycle laser fields: Frontiers of nonlinear optics},
  author = {Brabec, Thomas and Krausz, Ferenc},
  journal = {Rev. Mod. Phys.},
  volume = {72},
  issue = {2},
  pages = {545--591},
  numpages = {0},
  year = {2000},
  month = {Apr},
  publisher = {American Physical Society},
  doi = {10.1103/RevModPhys.72.545},
  url = {https://link.aps.org/doi/10.1103/RevModPhys.72.545}
}

@article{Phan:josab20,
author = {Ngoc-Loan Phan and Thanh-Tuynh Nguyen and Hirobumi Mineo and Van-Hung Hoang},
journal = {J. Opt. Soc. Am. B},
number = {2},
pages = {311--319},
publisher = {Optica Publishing Group},
title = {Depletion effect in high-order harmonic generation with coherent superposition state},
volume = {37},
month = {Feb},
year = {2020},
url = {https://opg.optica.org/josab/abstract.cfm?URI=josab-37-2-311},
doi = {10.1364/JOSAB.382154},
}

@article{Keldysh:JETP64,
  author       = {Keldysh, L V},
  title        = {IONIZATION IN THE FIELD OF A STRONG ELECTROMAGNETIC WAVE},
  url          = {https://www.osti.gov/biblio/4662394},
  journal      = {Zh. Eksperim. i Teor. Fiz. },
  volume       = {47},
  place        = {Country unknown/Code not available},
  year         = {1964},
  month        = {10}}

@article{Chirila:pra10,
  title={Emission times in high-order harmonic generation},
  author={Chiril{\u{a}}, CC and Dreissigacker, Ingo and van der Zwan, Elmar V and Lein, Manfred},
  journal={Phys. Rev. A},
  volume={81},
  number={3},
  pages={033412},
  year={2010},
  publisher={APS}
}

@article{Le:pra24,
  title = {Electron dynamical phase difference and high-order harmonic frequency shift: A {B}ohmian-trajectory perspective},
  author = {Le, Cam-Tu and Phan, Ngoc-Loan and Le, Van-Hoang},
  journal = {Phys. Rev. A},
  volume = {110},
  issue = {6},
  pages = {063115},
  numpages = {12},
  year = {2024},
  month = {Dec},
  publisher = {American Physical Society},
  doi = {10.1103/PhysRevA.110.063115},
  url = {https://link.aps.org/doi/10.1103/PhysRevA.110.063115}
}

@article{Le:pra23,
  title = {Dynamic core-electron-polarization effect on the high-order harmonic generation process from a quantum-trajectory perspective},
  author = {Le, Cam-Tu and Ngo, Cong and Phan, Ngoc-Loan and Vu, DinhDuy and Le, Van-Hoang},
  journal = {Phys. Rev. A},
  volume = {107},
  issue = {4},
  pages = {043103},
  numpages = {11},
  year = {2023},
  month = {Apr},
  publisher = {American Physical Society},
  doi = {10.1103/PhysRevA.107.043103},
  url = {https://link.aps.org/doi/10.1103/PhysRevA.107.043103}
}

@article{Mohebbi:aplB,
  title={Generation of isolated sub-40-attosecond pulse with a multicycle chirped laser and a static electric field},
  author={Mohebbi, Masoud},
  journal={ Appl. Phys. B},
  volume={122},
  number={2},
  pages={39},
  year={2016},
  publisher={Springer}
}

@article{Chergui:NatRevPhys23,
  title={Progress and prospects in nonlinear extreme-ultraviolet and {X}-ray optics and spectroscopy},
  author={Chergui, Majed and Beye, Martin and Mukamel, Shaul and Svetina, Cristian and Masciovecchio, Claudio},
  journal={Nat. Rev. Phys.},
  volume={5},
  number={10},
  pages={578--596},
  year={2023},
  publisher={Nature Publishing Group UK London}
}

@article{Paulus:jpb94,
doi = {10.1088/0953-4075/27/21/003},
url = {https://doi.org/10.1088/0953-4075/27/21/003},
year = {1994},
month = {nov},
publisher = {},
volume = {27},
number = {21},
pages = {L703},
author = {G G Paulus and W Becker and W Nicklich and H Walther},
title = {Rescattering effects in above-threshold ionization: a classical model},
journal = {J. Phys. B: At. Mol. Opt. Phys.}
}

@article{Majorosi:pra18,
  title = {Improved one-dimensional model potentials for strong-field simulations},
  author = {Majorosi, Szil\'ard and Benedict, Mih\'aly G. and Czirj\'ak, Attila},
  journal = {Phys. Rev. A},
  volume = {98},
  issue = {2},
  pages = {023401},
  numpages = {12},
  year = {2018},
  month = {Aug},
  publisher = {American Physical Society},
  doi = {10.1103/PhysRevA.98.023401},
  url = {https://link.aps.org/doi/10.1103/PhysRevA.98.023401}
}

@article{Li:pra20,
  title={Terahertz-field-induced near-cutoff even-order harmonics in a femtosecond laser},
  author={Li, Bing-Yu and Zhang, Jincang and Zhang, Yizhu and Yan, Tian-Min and Jiang, YH},
  journal={Phys. Rev. A},
  volume={102},
  number={6},
  pages={063102},
  year={2020},
  publisher={APS}
}

@article{Trieu2:pra24,
  title={Laser-target symmetry breaking in high-order harmonic generation: From frequency shift to odd-even intensity modulation},
  author={Trieu, Doan-An and Le, Van-Hoang and Phan, Ngoc-Loan},
  journal={Phys. Rev. A},
  volume={110},
  number={4},
  pages={L041101},
  year={2024},
  publisher={APS}
}

@article{Trieu:pra24,
  title={Analytically controlling the laser-induced electron phase in sub-cycle motion},
  author={Trieu, Doan-An and Nguyen, Trong-Thanh D and Nguyen, Thanh-Duy D and Tran, Thanh and Le, Van-Hoang and Phan, Ngoc-Loan},
  journal={Phys. Rev. A},
  volume={110},
  number={2},
  pages={L021101},
  year={2024},
  publisher={APS}
}

@article{Trieu:pra23,
  title={Universality in odd-even harmonic generation and application in terahertz waveform sampling},
  author={Trieu, Doan-An and Phan, Ngoc-Loan and Truong, Quan-Hao and Nguyen, Hien T and Le, Cam-Tu and Vu, DinhDuy and Le, Van-Hoang},
  journal={Phys. Rev. A},
  volume={108},
  number={2},
  pages={023109},
  year={2023},
  publisher={APS}
}

@article{Liu:pra25,
  title={Terahertz-field-modulated high-order harmonic generation in monolayer {M}o{S}$_2$},
  author={Liu, Xiulan and Zhang, Jianing and Fang, Yong-Kang and Yuan, Guanglu and Li, Zhengliang and Peng, Liang-You},
  journal={Phys. Rev. A},
  volume={111},
  number={6},
  pages={063103},
  year={2025},
  publisher={APS}
}

@article{Shulyndin:JETP25,
  title={Influence of the Terahertz Field on the Processes of Low-and High-Order Harmonic Generation by Femtosecond Laser Pulses in a Gaseous Medium},
  author={Shulyndin, Pavel Aleksandrovich and Rumiantsev, BV and Migal, Ekaterina Aleksandrovna and Pushkin, Andrei Vladimirovich and Potemkin, Fedor Vyktorovich},
  journal={JETP Lett.},
  volume={121},
  number={11},
  pages={846--852},
  year={2025},
  publisher={Springer}
}

@Article{Choi:JMCC24,
author ="Choi, Won Jin and Armstrong, Michael R. and Yoo, Jae Hyuck and Lee, Taeil",
title  ="Toward high-power terahertz radiation sources based on ultrafast lasers",
journal  ="J. Mater. Chem. C",
year  ="2024",
volume  ="12",
issue  ="25",
pages  ="9002-9011",
publisher  ="The Royal Society of Chemistry",
doi  ="10.1039/D4TC01502A",
url  ="http://dx.doi.org/10.1039/D4TC01502A"}

@article{Fulop:ThzRev20,
author = {Fülöp, József András and Tzortzakis, Stelios and Kampfrath, Tobias},
title = {Laser-Driven Strong-Field Terahertz Sources},
journal = {Adv. Opt. Mater.},
volume = {8},
number = {3},
pages = {1900681},
doi = {https://doi.org/10.1002/adom.201900681},
url = {https://advanced.onlinelibrary.wiley.com/doi/abs/10.1002/adom.201900681},
year = {2020}
}

@article{Rumiantsev:pra25,
  title={Observation of terahertz-field-induced coherent control of high-order harmonic generation in a noble gas},
  author={Rumiantsev, BV and Migal, EA and Pushkin, AV and Potemkin, FV},
  journal={Phys. Rev. A},
  volume={111},
  number={2},
  pages={023117},
  year={2025},
  publisher={APS}
}

@article{Taranukhin:josab00,
author = {Vladimir D. Taranukhin and Nickolay Yu. Shubin},
journal = {J. Opt. Soc. Am. B},
number = {9},
pages = {1509--1516},
publisher = {Optica Publishing Group},
title = {High-order harmonic generation by atoms with strong high-frequency and low-frequency pumping},
volume = {17},
month = {Sep},
year = {2000},
url = {https://opg.optica.org/josab/abstract.cfm?URI=josab-17-9-1509},
doi = {10.1364/JOSAB.17.001509},
}

@article{Raab:RevSciInstru24,
    author = {Raab, A.-K. and Schmoll, M. and Simpson, E. R. and Redon, M. and Fang, Y. and Guo, C. and Viotti, A.-L. and Arnold, C. L. and L’Huillier, A. and Mauritsson, J.},
    title = {Highly versatile, two-color setup for high-order harmonic generation using spatial light modulators},
    journal = {Rev. Sci. Instrum.},
    volume = {95},
    number = {7},
    pages = {073002},
    year = {2024},
    month = {07},
    issn = {0034-6748},
    doi = {10.1063/5.0212578},
}

@article{Luu:NatCom18,
  title={Extreme--ultraviolet high--harmonic generation in liquids},
  author={Luu, Tran Trung and Yin, Zhong and Jain, Arohi and Gaumnitz, Thomas and Pertot, Yoann and Ma, Jun and W{\"o}rner, Hans Jakob},
  journal={ Nat. Commun.},
  volume={9},
  number={1},
  pages={3723},
  year={2018},
  publisher={Nature Publishing Group UK London}
}

@article{Vampa:jpb17,
doi = {10.1088/1361-6455/aa528d},
url = {https://doi.org/10.1088/1361-6455/aa528d},
year = {2017},
month = {mar},
publisher = {IOP Publishing},
volume = {50},
number = {8},
pages = {083001},
author = {Vampa, G and Brabec, T},
title = {Merge of high harmonic generation from gases and solids and its implications for attosecond science},
journal = {J. Phys. B: At. Mol. Opt. Phys.}
}

@article{Lewenstein:pra94,
  title = {Theory of high-harmonic generation by low-frequency laser fields},
  author = {Lewenstein, M. and Balcou, Ph. and Ivanov, M. Yu. and L'Huillier, Anne and Corkum, P. B.},
  journal = {Phys. Rev. A},
  volume = {49},
  issue = {3},
  pages = {2117--2132},
  numpages = {0},
  year = {1994},
  month = {Mar},
  publisher = {American Physical Society},
  doi = {10.1103/PhysRevA.49.2117},
  url = {https://link.aps.org/doi/10.1103/PhysRevA.49.2117}
}

@article{Corkum:prl93,
  title = {Plasma perspective on strong field multiphoton ionization},
  author = {Corkum, P. B.},
  journal = {Phys. Rev. Lett.},
  volume = {71},
  issue = {13},
  pages = {1994--1997},
  numpages = {0},
  year = {1993},
  month = {Sep},
  publisher = {American Physical Society},
  doi = {10.1103/PhysRevLett.71.1994},
  url = {https://link.aps.org/doi/10.1103/PhysRevLett.71.1994}
}

@article{Borrego-Varillas:RepProPhys22,
doi = {10.1088/1361-6633/ac5e7f},
url = {https://doi.org/10.1088/1361-6633/ac5e7f},
year = {2022},
month = {may},
publisher = {IOP Publishing},
volume = {85},
number = {6},
pages = {066401},
author = {Borrego-Varillas, Rocío and Lucchini, Matteo and Nisoli, Mauro},
title = {Attosecond spectroscopy for the investigation of ultrafast dynamics in atomic, molecular and solid-state physics},
journal = {Rep. Prog. Phys.}
}

@article{Weissenbilder:NatRevPhys22,
  title={How to optimize high-order harmonic generation in gases},
  author={Weissenbilder, R and Carlstr{\"o}m, S and Rego, L and Guo, C and Heyl, CM and Smorenburg, P and Constant, Eric and Arnold, CL and L’Huillier, A},
  journal={Nat. Rev. Phys.},
  volume={4},
  number={11},
  pages={713--722},
  year={2022},
  publisher={Nature Publishing Group UK London}
}

@article{Popmintchev:pnas09,
author = {Tenio Popmintchev  and Ming-Chang Chen  and Alon Bahabad  and Michael Gerrity  and Pavel Sidorenko  and Oren Cohen  and Ivan P. Christov  and Margaret M. Murnane  and Henry C. Kapteyn },
title = {Phase matching of high harmonic generation in the soft and hard X-ray regions of the spectrum},
journal = {Proc. Natl. Acad. Sci.},
volume = {106},
number = {26},
pages = {10516-10521},
year = {2009},
URL = {https://www.pnas.org/doi/abs/10.1073/pnas.0903748106}}

@article{Hong:OE09,
  title={Few-cycle attosecond pulses with stabilized-carrier-envelope phase in the presence of a strong terahertz field},
  author={Hong, Weiyi and Lu, Peixiang and Lan, Pengfei and Zhang, Qingbin and Wang, Xinbing},
  journal={Opt. Express},
  volume={17},
  number={7},
  pages={5139--5146},
  year={2009},
  publisher={Optica Publishing Group}
}

@article{Ge:OE15,
  title={Quantum control of electron wave packet during high harmonic process of {H}$_2^+$ in a combination of a circularly polarized laser field and a Terahertz field},
  author={Ge, Xin-Lei and Du, Hui and Guo, Jing and Liu, Xue-Shen},
  journal={Opt. Express},
  volume={23},
  number={7},
  pages={8837--8844},
  year={2015},
  publisher={Optica Publishing Group}
}

@article{Wang:JPB98,
  title={The effects of a static electric field on high-order harmonic generation},
  author={Wang, Bingbing and Li, Xiaofeng and Fu, Panming},
  journal={J. Phys. B: At. Mol. Opt. Phys.},
  volume={31},
  number={9},
  pages={1961},
  year={1998},
  publisher={IOP Publishing}
}

@article{Hong:JPB07,
  title={Control of quantum paths of high-order harmonics and attosecond pulse generation in the presence of a static electric field},
  author={Hong, Weiyi and Lu, Peixiang and Cao, Wei and Lan, Pengfei and Wang, Xinlin},
  journal={J. Phys. B: At. Mol. Opt. Phys.},
  volume={40},
  number={12},
  pages={2321},
  year={2007},
  publisher={IOP Publishing}
}

@article{Koushki:cp21,
  title={High-order harmonic generation from CH4 and CD4 molecules in the presence of a static electric field},
  author={Koushki, AM and Sarikhani, S},
  journal={Chem. Phys.},
  volume={541},
  pages={111020},
  year={2021},
  publisher={Elsevier}
}

@article{Odzak:pra05,
  title = {High-order harmonic generation in the presence of a static electric field},
  author = {Od\ifmmode \check{z}\else \v{z}\fi{}ak, S. and Milo\ifmmode \check{s}\else \v{s}\fi{}evi\ifmmode \acute{c}\else \'{c}\fi{}, D. B.},
  journal = {Phys. Rev. A},
  volume = {72},
  issue = {3},
  pages = {033407},
  numpages = {9},
  year = {2005},
  month = {Sep},
  publisher = {American Physical Society},
  doi = {10.1103/PhysRevA.72.033407},
  url = {https://link.aps.org/doi/10.1103/PhysRevA.72.033407}
}

@article{Bandrauk:PRA97,
  title={Enhanced harmonic generation in extended molecular systems by two-color excitation},
  author={Bandrauk, AD and Chelkowski, S and Yu, H and Constant, E},
  journal={Phys. Rev. A},
  volume={56},
  number={4},
  pages={R2537},
  year={1997},
  publisher={APS}
}

@article{Yuan:pra11,
  title = {Circularly polarized molecular high-order harmonic generation in ${{\mathrm{H}}_{2}}^{+}$ with intense laser pulses and static fields},
  author = {Yuan, Kai-Jun and Bandrauk, Andr\'e D.},
  journal = {Phys. Rev. A},
  volume = {83},
  issue = {6},
  pages = {063422},
  numpages = {7},
  year = {2011},
  month = {Jun},
  publisher = {American Physical Society},
  doi = {10.1103/PhysRevA.83.063422},
  url = {https://link.aps.org/doi/10.1103/PhysRevA.83.063422}
}

@article{Li:NC23,
  title={High-order harmonic generation from a thin film crystal perturbed by a quasi-static terahertz field},
  author={Li, Sha and Tang, Yaguo and Ortmann, Lisa and Talbert, Bradford K and Blaga, Cosmin I and Lai, Yu Hang and Wang, Zhou and Cheng, Yang and Yang, Fengyuan and Landsman, Alexandra S and Agostini, Pierre and DiMauro, Louis F.},
  journal={Nat. Commun.},
  volume={14},
  number={1},
  pages={2603},
  year={2023},
  publisher={Nature Publishing Group UK London}
}

@article{Carrera:PRA07,
  title = {Extension of high-order harmonic generation cutoff via coherent control of intense few-cycle chirped laser pulses},
  author = {Carrera, Juan J. and Chu, Shih-I},
  journal = {Phys. Rev. A},
  volume = {75},
  issue = {3},
  pages = {033807},
  numpages = {5},
  year = {2007},
  month = {Mar},
  publisher = {American Physical Society},
  doi = {10.1103/PhysRevA.75.033807},
  url = {https://link.aps.org/doi/10.1103/PhysRevA.75.033807}
}

@article{Ge:LPL20,
  title={The comparison of high-order harmonic generation of H $ \_ $\{$2$\}$\^{}$\{$+$\}$ $ in a circularly polarized laser pulse combined with a terahertz field and with a static electric field},
  author={Ge, Xin-Lei},
  journal={Laser Phys. Lett.},
  volume={17},
  number={5},
  pages={055301},
  year={2020},
  publisher={IOP Publishing}
}

\end{document}